\documentclass[aps,prd,twocolumn,preprintnumbers,superscriptaddress,floatfix,nofootinbib,notitlepage,showkeys,showpacs]{revtex4-1}

\usepackage[utf8x]{inputenc}
\usepackage{graphicx}
\usepackage{hyperref}
\usepackage{amsmath}
\usepackage{amssymb}
\usepackage{color}
\usepackage{enumitem}

\definecolor{lmpurple}{RGB}{151, 9, 222}

\def\lsim{\mathrel{\raise.3ex\hbox{$<$\kern-.75em\lower1ex\hbox{$\sim$}}}}
\def\gsim{\mathrel{\raise.3ex\hbox{$>$\kern-.75em\lower1ex\hbox{$\sim$}}}}

\begin{document}
\rightline{CERN-TH-2018-072, KCL-PH-TH/2018-13}
\vspace{0.5cm}
\title{Dark matter effects on neutron star properties}
\author{John Ellis}
\email{john.ellis@cern.ch}
\affiliation{NICPB, R\"avala pst.~10, 10143 Tallinn, Estonia}
\affiliation{ Theoretical Particle Physics and Cosmology Group, Physics Department, King's College London, London WC2R 2LS, United Kingdom}
\affiliation{Theoretical Physics Department, CERN, CH-1211 Geneva 23, Switzerland}
\author{Gert H{\"u}tsi}
\email{gert.hutsi@to.ee}
\affiliation{NICPB, R\"avala pst.~10, 10143 Tallinn, Estonia}
\affiliation{Tartu Observatory, T{\~o}ravere 61602, Estonia}
\author{Kristjan Kannike}
\email{kristjan.kannike@cern.ch}
\affiliation{NICPB, R\"avala pst.~10, 10143 Tallinn, Estonia}
\author{Luca Marzola}
\email{luca.marzola@cern.ch}
\affiliation{NICPB, R\"avala pst.~10, 10143 Tallinn, Estonia}
\author{Martti Raidal}
\email{martti.raidal@cern.ch}
\affiliation{NICPB, R\"avala pst.~10, 10143 Tallinn, Estonia}
\author{Ville Vaskonen}
\email{ville.vaskonen@kbfi.ee}
\affiliation{NICPB, R\"avala pst.~10, 10143 Tallinn, Estonia}
\begin{abstract}
We study possible effects of a dark matter (DM) core on the maximum mass of a neutron star (NS), on the mass-radius relation and on the NS tidal deformability parameter $\Lambda$. We show that all these quantities would in general be reduced in the presence of a DM core. In particular, our calculations indicate that the presence of a DM core with a mass fraction $\sim 5\%$ could affect significantly the interpretation of these NS data as constraints on the nuclear equation of state (EOS), potentially excluding some EOS models on the basis of the measured mass of PSR J0348+0432, while allowing other EOS models to become consistent with the LIGO/Virgo upper limit on $\Lambda$. Specific scenarios for generating such DM cores are explored in an Appendix.
\end{abstract}

\maketitle

%-------------------------------------------------------------------------------
\section{Introduction}
%-------------------------------------------------------------------------------

The discovery of gravitational waves (GWs) by the Advanced LIGO and Virgo Collaborations~\cite{Abbott:2016blz} has opened
a new window on the Universe, making possible novel probes of gravitational physics, astrophysics,
cosmology and other other aspects of fundamental physics. Moreover, the observation of GWs from
the merger of a pair of neutron stars (NSs)~\cite{TheLIGOScientific:2017qsa}, by constraining the tidal deformability $\Lambda$ of dense 
nuclear matter, has provided a new probe of the nuclear equation of state (EOS)~\cite{Annala:2017llu}.
We pointed out~\cite{Ellis:2017jgp} that this observation of GWs from a NS-NS merger also opened a new window on 
possible models of dark matter (DM) that could modify the GW signal emitted following the
merger, yielding one or two additional peaks in the postmerger frequency spectrum 
that might be detectable in the future GW signals from NS mergers.
Previous to our work, it was also pointed out that DM could have a significant effect on the NS mass-radius 
relation~\cite{Panotopoulos:2017idn}, and more recently the possible effect of DM on 
the tidal deformability of a NS has been considered~\cite{Nelson:2018xtr}.

The DM scenario considered in~\cite{Ellis:2017jgp} invoked the presence of a DM core inside the NS, whereas that envisaged in~\cite{Nelson:2018xtr} was of a DM halo enveloping the star. In this paper we extend the study of our DM core scenario to include considerations of its possible effects on the maximum mass of a NS, its radius for any fixed mass, and its tidal deformability $\Lambda$. All of these quantities are in general {\it reduced} in the presence of such a DM core. For comparison, we note that reductions in the maximum mass and radius were also found in the DM core model of~\cite{Panotopoulos:2017idn}, whereas an increase in $\Lambda$ was found in the DM halo model of~\cite{Nelson:2018xtr}. Within the framework of the DM core model we consider, the observation of the neutron star PSR J0348+0432~\cite{Antoniadis:2013pzd}, with a mass $\simeq 2 \, M_\odot$, limits the choice of possible EOSs, whereas a larger selection of EOSs would be compatible with the LIGO/Virgo upper limit $\Lambda < 800$ on the tidal deformability, if the DM core has a mass of 5\% of the total NS mass. In this case, for example, the EOS ALF2, H4 and SLy4 would be ruled out by PSR J0348+0432, whereas the EOS APR3, MS1, MS1B and H4 could no longer be excluded by the  upper limit on $\Lambda$. (See~\cite{Engvik:1995gn,Potekhin:2013qqa,Muther:1987xaa,Lackey:2005tk,Alford:2004pf,Akmal:1998cf,Douchin:2001sv,Mueller:1996pm} for references to these models. In Ref.~\cite{Akmal:1998cf} the EOS APR4 and APR3 are called A18+$\delta{v}$+UIX$\mbox{}^*$ and A18+UIX, respectively.)

The layout of this paper is as follows. In Section~\ref{sec:model} we discuss how we model the nuclear and
DM components of the NS. For the former we consider 11 representative EOSs taken from~\cite{Engvik:1995gn,Potekhin:2013qqa,Muther:1987xaa,Lackey:2005tk,Alford:2004pf,Akmal:1998cf,Douchin:2001sv,Mueller:1996pm} that are compatible with
the observation of PSR J0348+0432~\cite{Antoniadis:2013pzd}, and for the DM EOS we consider in detail an
example appropriate for self-interacting bosonic DM, arguing that asymmetric DM with or without self-interactions
would yield similar results. In Section~\ref{sec:MRrelation} we calculate for various EOS the effect on the NS mass-radius
relation of a DM core with 5\% of the NS mass. We also explore in more detail the effects of
the DM self-interaction strength and core mass for the H4 EOS~\cite{Lackey:2005tk}, which is chosen because
it has large maximum mass, considering also the possibility 
that the DM forms a halo with radius larger than the nuclear matter. The possible effects of a DM core on the tidal
deformability are considered in Section~\ref{sec:tidaldeformability}, and our conclusions are summarized in
Section~\ref{sec:conclusions}. Finally, for completeness we mention in an Appendix three possible scenarios for the 
formation of the DM cores and halos that we consider in the body of the paper, one invoking $n \to$ massive
DM conversion, another postulating bremsstrahlung of a lighter DM particle, and a third based on the existence of DM stars.

%-------------------------------------------------------------------------------
\section{Modeling the neutron star components}
\label{sec:model}
%-------------------------------------------------------------------------------

In order to be consistent with the observation of PSR J0348+0432~\cite{Antoniadis:2013pzd}, we
use 11 different nuclear EOSs that reproduce a maximal NS mass exceeding $1.97M_\odot$
to model the baryonic component of a NS~\cite{Engvik:1995gn,Potekhin:2013qqa,Muther:1987xaa,Lackey:2005tk,Alford:2004pf,Akmal:1998cf,Douchin:2001sv,Mueller:1996pm}. As illustrations of the possible effects of a DM component
we have considered the following possibilities.
\begin{itemize}[leftmargin=*]
	\item {\it Self-interacting bosonic DM}: We use the Bose-Einstein condensate EOS~\cite{Li:2012qf},  
\begin{equation} \label{eq:DMEOS}
P_{\rm D} = \frac{\hbar^2\sqrt{\pi\sigma_{\rm D}}}{m_{\rm D}^3} \rho_{\rm D}^2\,,
\end{equation}
where $\sigma_{\rm D}$ is the repulsive DM self-interaction cross section and $m_{\rm D}$ is the mass of the DM particles. 

\item {\it Asymmetric fermionic DM}: In the absence of self-interactions, the pressure and energy density are
given, respectively, by~\cite{Kouvaris:2015rea,Mukhopadhyay:2016dsg}
\begin{equation}
	\label{eq:fdmni}
\rho_{\rm D}^{(0)} = \frac{m_{\rm D}^4 c^6}{\hbar^3} \chi(x) \,,\quad P_{\rm D}^{(0)} = \frac{m_{\rm D}^4 c^6}{\hbar^3} \phi(x) \,,
\end{equation}
where
\begin{equation}
\begin{aligned}
\chi(x) =  \frac{x\sqrt{1+x^2}(1+2x^2) - \ln\left(x+ \sqrt{1+x^2}\right)}{8\pi^2} \,,\\
\phi(x) = \frac{x\sqrt{1+x^2}\left(\frac{2x^2}{3}-1\right) + \ln\left(x+ \sqrt{1+x^2}\right)}{8\pi^2} \,,
\end{aligned}
\end{equation} 
and the parameter $x=p_{\rm D}/m_{\rm D}$ quantifies how relativistic the DM particles are, where $p_{\rm D}$
denotes the Fermi momentum. Including now a self-interaction potential
\begin{equation}
V = \frac{g \,e^{-m_a r}}{4\pi r}\,,
\end{equation}
where $g$ is the DM-mediator coupling constant and $m_a$ the mediator mass, the quantities in Eq.~\eqref{eq:fdmni} receive
the following additional terms~\cite{Kouvaris:2015rea,Mukhopadhyay:2016dsg}:
\begin{equation}
\begin{aligned}
&\rho_{\rm D} = \rho_{\rm D}^{(0)} + \frac{g^2x^6m_{\rm D}^6}{2(3\pi^2)^2 (\hbar c)^3 m_a^2} \,, \\ 
&P_{\rm D} = P_{\rm D}^{(0)} + \frac{g^2x^6m_{\rm D}^6}{2(3\pi^2)^2 (\hbar c)^3 m_a^2} \,.
\end{aligned}
\end{equation} 
\end{itemize}

We anticipate that, once a positive pressure is generated for the DM component so as to prevent the formation of a black hole at the center of the
NS~\cite{Bertone:2007ae,deLavallaz:2010wp,Kouvaris:2011fi,McDermott:2011jp}, these examples
would result in qualitatively similar phenomenological effects on the mass-radius relation and the tidal deformability of NSs. 
As we will show in the next sections, these observables are insensitive to the dynamics that stabilize the DM structure, be it the Fermi pressure
or the repulsive self-interactions necessary in the bosonic case, and only track the total DM mass.  In the following we
present a detailed analysis of the effects of a DM component modeled using the EOS in Eq.~\eqref{eq:DMEOS}, 
expecting that similar results would hold in the other cases.

%-------------------------------------------------------------------------------
\section{Mass-Radius Relation for Neutron Stars}
\label{sec:MRrelation}
%-------------------------------------------------------------------------------

\begin{figure}[t]
\centering
\includegraphics[height=.73\linewidth]{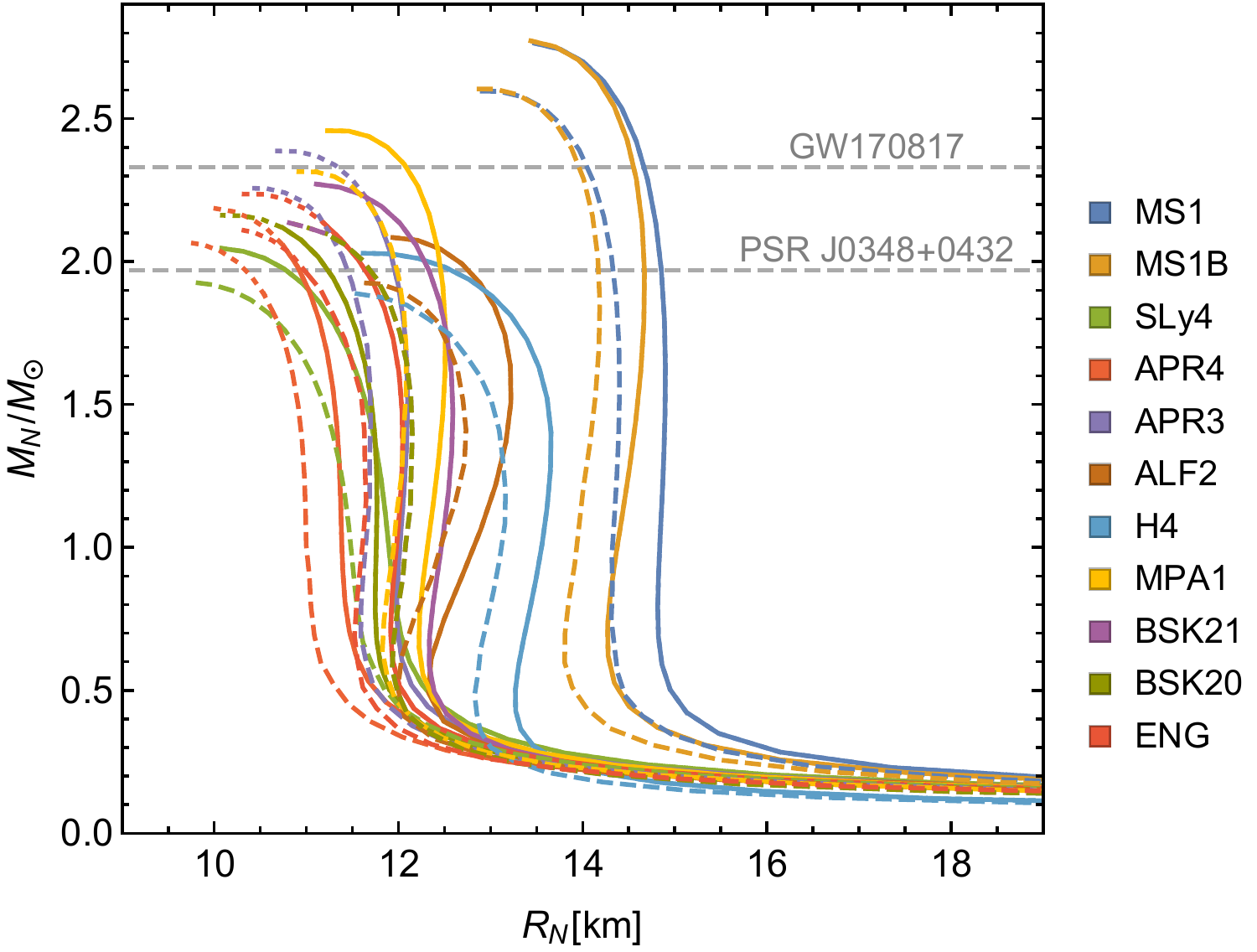} 
\caption{Effects of DM cores on the mass-radius relation for NSs. The color code denotes the different baryonic EOSs considered~\cite{Engvik:1995gn,Potekhin:2013qqa,Muther:1987xaa,Lackey:2005tk,Alford:2004pf,Akmal:1998cf,Douchin:2001sv,Mueller:1996pm}. The solid lines are for cases in which a NS contains no DM, whereas the dashed lines include the effect of a DM core contributing to 5\% of the total NS mass and $\sqrt{\sigma_{\rm D}}/m_{\rm D}^3 = 0.05\,{\rm GeV}^{-2}$. The horizontal lines show the bounds on the maximal NS mass $1.97 M_\odot < M_{\rm max} < 2.33 M_\odot$ inferred, respectively, by observations of PSR J0348+0432~\cite{Antoniadis:2013pzd} and an analysis of GW170817~\cite{Rezzolla:2017aly}, assuming that its merger product was a black hole.}
\label{fig:RM5}
\end{figure}

\begin{figure*}[t]
\centering
\includegraphics[height=.34\linewidth]{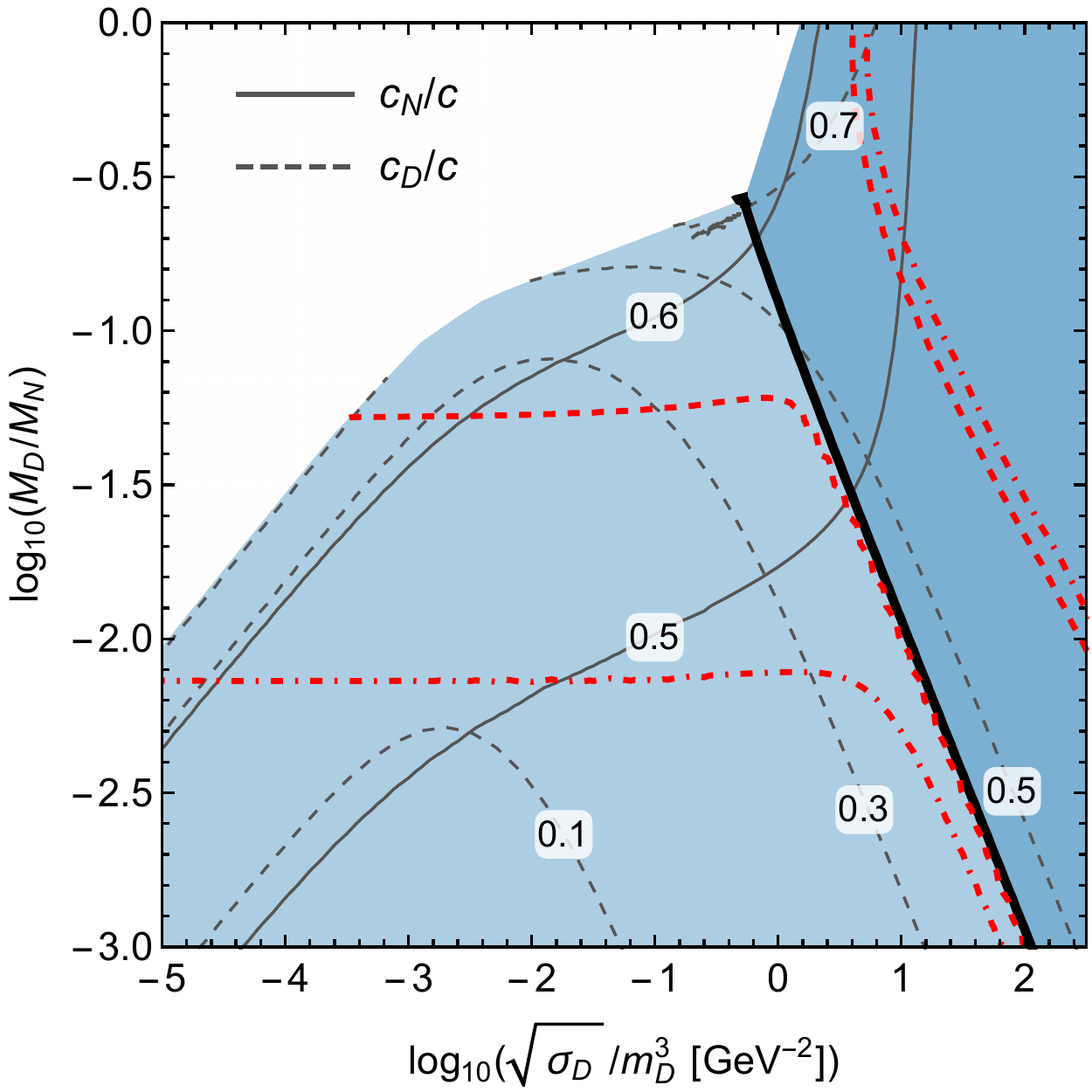} \hspace{-2mm}
\includegraphics[height=.34\linewidth]{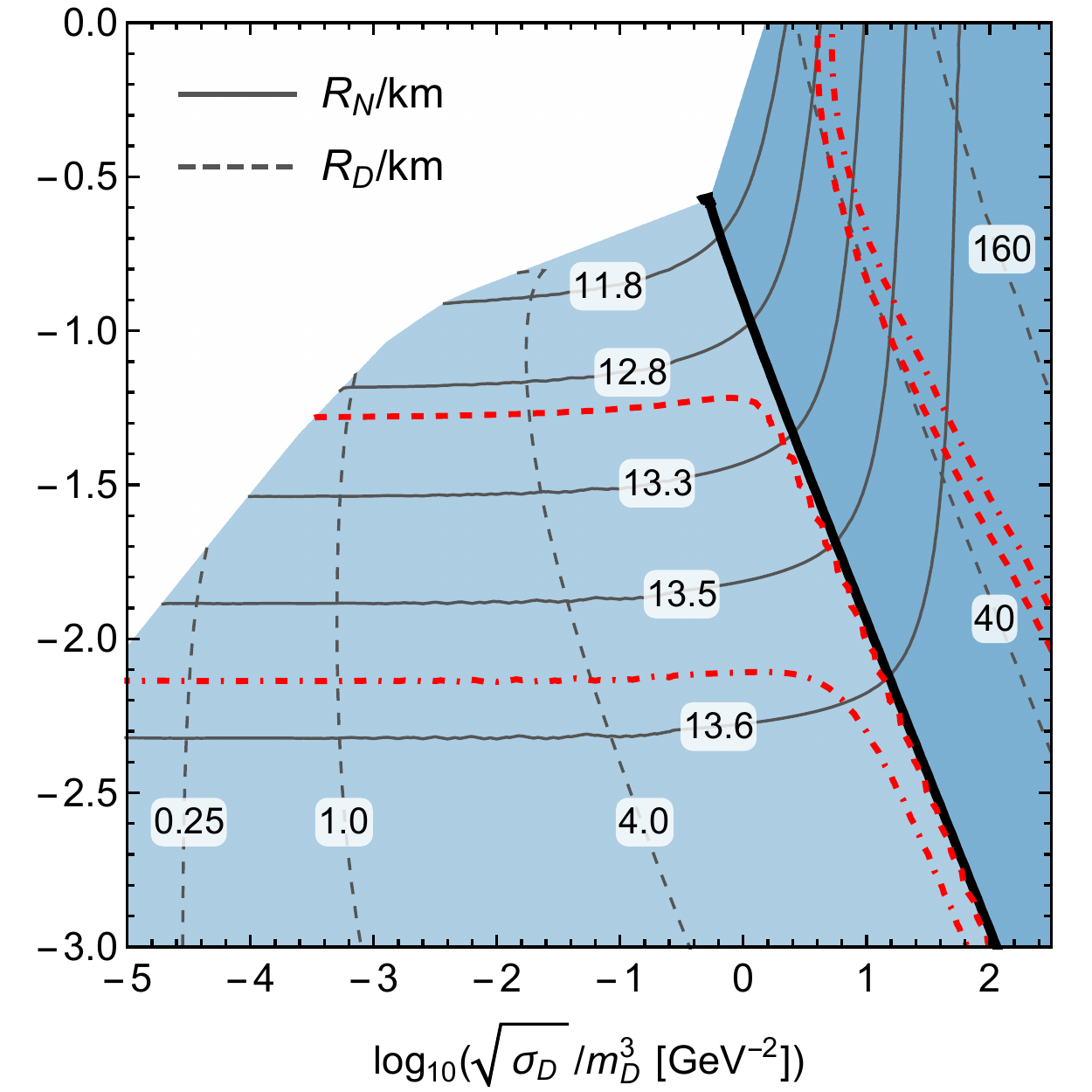} \hspace{-2mm}
\includegraphics[height=.34\linewidth]{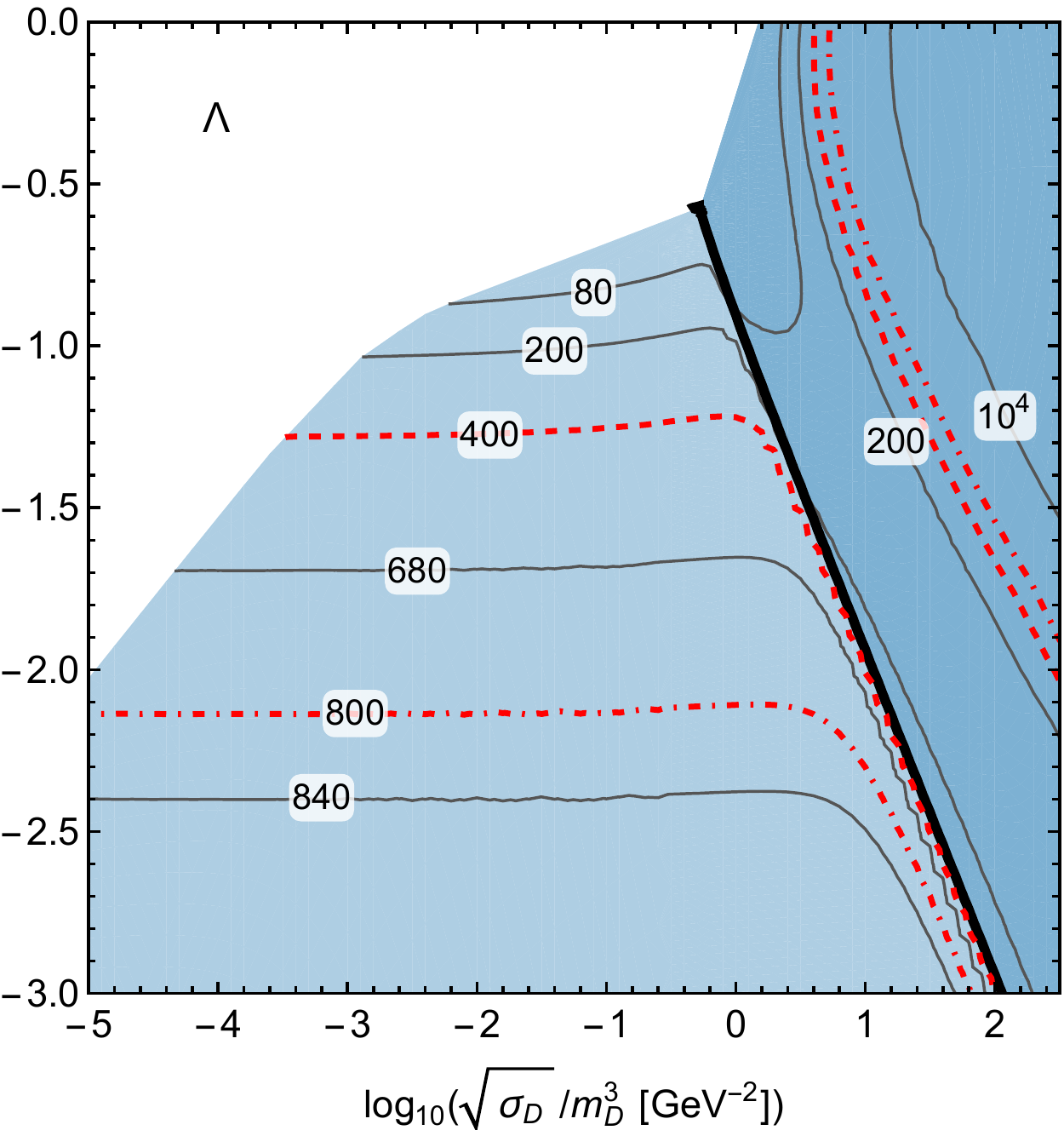}
\caption{The lighter regions show the parameter space where DM forms a core inside the NS,
assuming the H4 EOS for the nuclear component. In the darker regions, to the contrary, 
the DM halo envelops a baryonic NS core. White areas are excluded by the stability constraint. All the 
plots are for a baryonic mass of $M_{\rm N} = 1.4M_\odot$. The left panel 
shows the sound velocities $c_{\rm N}$ and $c_{\rm D}$ in the baryonic and DM components, respectively, at the center of the NS. 
The middle panel shows the radii $R_{\rm N}$ and $R_{\rm D}$ of the baryonic and DM components, respectively, and the right 
panel shows the tidal deformability $\Lambda$. In each panel {the red dashed and dot-dashed curves show the contours 
$\Lambda=400$, and $\Lambda=800$, respectively.}}
\label{fig:DMF}
\end{figure*}

We have solved the coupled Tolman-Oppenheimer-Volkoff (TOV) equations~\cite{Oppenheimer:1939ne,Tolman:1934za}
to compute the radial density and pressure profiles of the baryonic and DM components
for various choices of the nuclear EOS and ranges of the effective DM self-interaction strength  
$\sqrt{\sigma_{\rm D}}/m_{\rm D}^3$ and the DM fraction. We assume that the nongravitational interactions 
between the nuclear and DM components are negligible, so that they interact only via the 
common  gravitational potential~\cite{Sandin:2008db}. 

The effect of a DM core on the mass-radius relation is shown in Fig.~\ref{fig:RM5}. Here the color code denotes the different 
EOSs considered, with solid and dashed lines representing the cases in which the DM core is absent or contributes
5\% of the total NS mass, respectively. The ends of the lines at low $R$ are due to instability of the 
NS~\cite{Shapiro:1983du}. Dotted lines indicate instead that the causality condition~\cite{Haensel:2007yy} is not satisfied. 
The overall effect of a DM core is to make the NS more compact and to decrease the maximum NS mass achievable.
The reduction in the maximum mass for a DM fraction of 5\% would render the EOS ALF2, APR4, ENG, H4 and SLy4 incompatible
with the measurement of the mass of PSR J0348+0432. 

The upper dashed horizontal line in Fig.~\ref{fig:RM5} shows the upper bound on the maximal NS mass obtained in 
Ref.~\cite{Rezzolla:2017aly} by combining the observations of the NS merger GW170817 with a quasiuniversal relation 
between the maximum mass of nonrotating and uniformly rotating stellar models, and assuming that the merger product 
collapses to a black hole. As a DM core would reduce the maximum 
mass for a given baryonic EOS, its presence would increase increase the compatibility of any EOS
with this bound. Moreover, we expect this bound to change in the presence of a DM core within the merging NSs. 
For instance, the bound would soften if a substantial DM core remains inside the merger product. 

Since the H4 nuclear EOS is in good agreement with the bounds mentioned above, we explore in more detail 
in Fig.~\ref{fig:DMF} the effects of the DM self-interaction strength and core mass for this EOS.
In the left and middle panels of Fig.~\ref{fig:DMF} we show the dependences of the NS radius and
the sound speed at the center of the NS on the DM self-interaction strength and total DM core mass. 
These calculations are for a NS with mass $M_{\rm N} = 1.4M_\odot$. In the shaded regions to the left of the thick solid lines 
the DM core is fully contained in the NS. We find that the radius of the NS does not depend significantly 
on the strength of the DM self-interaction. In the darker shaded regions to the right of the thick solid lines
the NS is enveloped within a DM halo. In this case also the presence of DM decreases the NS radius,
but now the latter is sensitive to the DM self-interaction strength. In essence, we find that the 
NS mass-radius relation is affected only by the total DM mass contained in the baryonic shell, 
regardless of its distribution. The NS stability requirement is not respected in the white regions of Fig.~\ref{fig:DMF}, 
and we see that this forbids DM cores with masses $\gtrsim 0.3 M_\odot$.

Similar results on the change of the compactness of the NS in the presence of a DM core have been obtained 
in Refs.~\cite{Sandin:2008db,Ciarcelluti:2010ji,Tolos:2015qra}. As pointed out in Ref.~\cite{Ciarcelluti:2010ji}, depending on the formation history 
of the DM core, the total DM mass contained in the NS may vary considerably. Consequently, we expect a spread in the compactness of NSs and the 
observed NSs need not obey the same mass-radius relation.

%-------------------------------------------------------------------------------
\section{Effect of DM on the Tidal Deformability}
\label{sec:tidaldeformability}
%-------------------------------------------------------------------------------
In the absence of a DM core, the gravitational radiation emitted during the inspiral of a NS binary system
is sensitive to the tidal distortion of the NS, which is sensitive in turn to the nuclear EOS~\cite{TheLIGOScientific:2017qsa}.
We now investigate the effect of a DM component on the dimensionless quadrupolar tidal parameter, defined as
\begin{equation}
\Lambda = \frac{2}{3} k_2 \left(\frac{GM}{c^2R}\right)^{-5}\,,
\label{TD}
\end{equation}
where $k_2$ is the $l = 2$ tidal Love number~\cite{Hinderer:2007mb,Postnikov:2010yn},
which can be calculated once the coupled TOV equations are solved for the total pressure and density profiles. 

\begin{figure}[t]
\centering
\includegraphics[height=.73\linewidth]{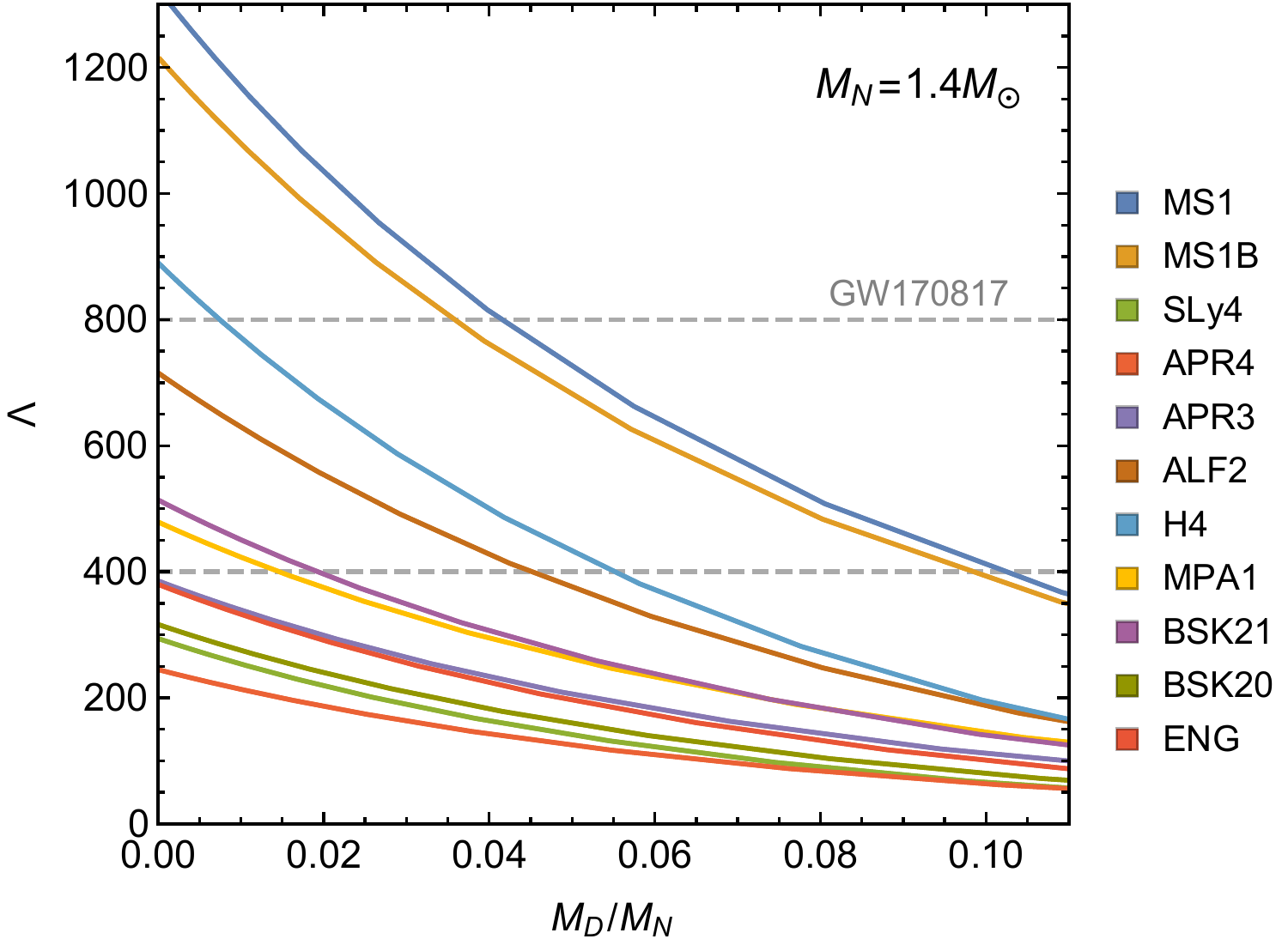}
\caption{The lines show how the tidal deformability $\Lambda$ (\protect\ref{TD}) changes as a function of the DM mass fraction,
with the DM self-interaction strength set to $\sqrt{\sigma_{\rm D}}/m_{\rm D}^3 = 0.05\,{\rm GeV}^{-2}$. 
Different colors represent different baryonic matter EOSs, as indicated. The total mass of the star is set to $1.4M_\odot$, 
within the range indicated by GW170817. Observations of its GW signal exclude the region above the upper 
dashed line at the $90\%$ confidence level~\cite{TheLIGOScientific:2017qsa}. The lower dashed line indicates a tentative lower 
bound on $\Lambda$~\cite{Radice:2017lry}.}
\label{fig:FL14}
\end{figure}

We have investigated the possible effect of DM on the tidal deformability $\Lambda$ (\ref{TD}), 
as seen in the right panel of Fig.~\ref{fig:DMF} and in Fig.~\ref{fig:FL14} for a fixed NS mass.
In the former we consider the effects of both the DM mass fraction and the self-interaction strength,
whereas in the latter we restrict our attention to small DM self-interactions, so that a DM core is formed rather than 
a halo surrounding the star. In both figures we take the NS mass to be $M_{\rm N}=1.4M_\odot$, which is consistent with the
recent measurements of the NS-NS merger GW170817~\cite{TheLIGOScientific:2017qsa}. 

As seen in Fig.~\ref{fig:FL14}, we find that, regardless of the baryonic EOS considered, 
the tidal deformability decreases progressively for increasing values of the DM core mass. 
The upper dashed line indicates the bound $\Lambda<800$ imposed by the GW observations at
the $90\%$ confidence level~\cite{TheLIGOScientific:2017qsa}. The possible presence of a
DM core could, therefore, complicate the interpretation of this upper bound, as well as possible
future measurements of NS-NS mergers. In particular, we see that in the presence of a DM core the
EOS MS1, MS1B and H4 could no longer be excluded by the upper bound on $\Lambda$
set by~\cite{TheLIGOScientific:2017qsa}. {In addition, the lower dashed line indicates a tentative 
bound $\Lambda > 400$ on the tidal deformability parameter obtained in Ref.~\cite{Radice:2017lry}  by 
combining optical/infrared and GW observations. If confirmed, this would rule out the EOS ENG, 
BSK20, SLy4, APR3 and APR4, and large DM cores for any baryonic EOS.}

More details of the dependence of the tidal deformability for $M_{\rm N} = 1.4M_\odot$ on the DM self-interaction strength and 
the total DM mass are shown in the right panels of Fig.~\ref{fig:DMF} for the nuclear EOS H4. 
Similarly to the NS radius, also the tidal deformability is essentially determined by the total DM mass contained within the 
baryonic core of the NS, and does not depend significantly on the DM self-interaction strength in the presence of a DM core. However, 
when the radius of the DM distribution is larger than the NS, the tidal deformability becomes sensitive to both the 
DM self-interaction strength and total mass. In the case that $R_{\rm D}\gg R_{\rm N}$, the tidal deformability increases 
for increasing values of $M_{\rm D}$, in contrast with the case of a DM core.  Such a behavior was to be expected, 
as the tidal deformability depends strongly on $M/R$, where $M$ and $R$ are the total mass and the total radius, respectively, of the 
combined NS + DM system, and the effect of DM on the tidal deformability is significant even when a very small total DM mass is considered.

Gravitational wave observations of the inspiral phase of a NS coalescence probe the last few dozens of seconds before the NS-NS
merger. During this phase, the separation of the stars is typically $r<150\,{\rm km}$~\cite{Nelson:2018xtr}, 
so observing the tidal deformability in the case $R_{\rm D}>R_{\rm N}$ requires that $R_{\rm D}\lsim 75\,{\rm km}$. 
We find that in this case the maximum values of the tidal deformability for $M_{\rm N} = 1.4M_\odot$ are $\Lambda = \mathcal{O}(10^3)$ 
for the H4 EOS. We note that this value is larger than the current $90\%$ confidence level bound 
$\Lambda<800$~\cite{TheLIGOScientific:2017qsa}, which is shown by the red dot-dashed lines in Fig.~\ref{fig:DMF}. 

\begin{figure}[t]
\centering
\includegraphics[width=.95\linewidth]{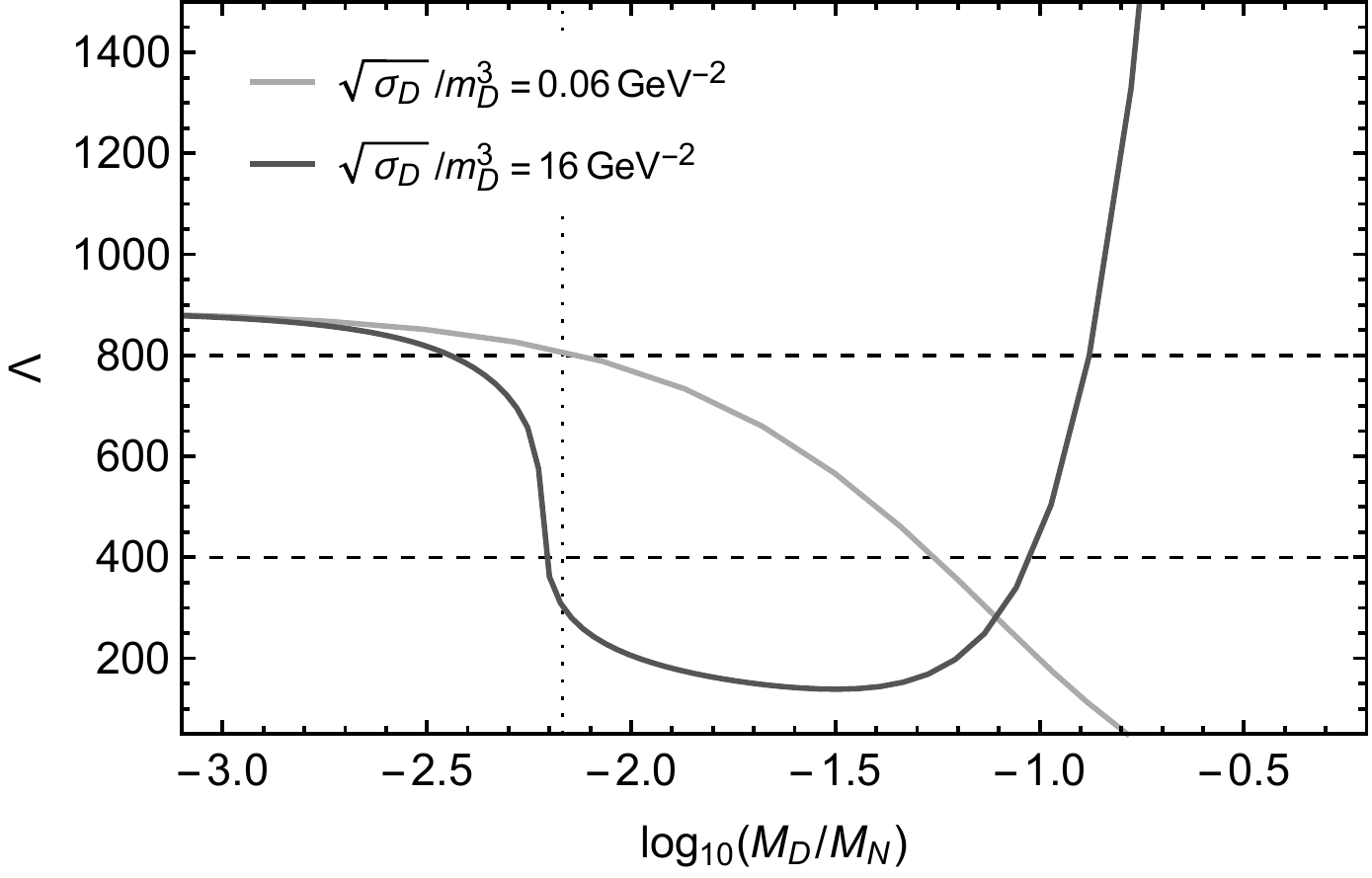}
\caption{The solid lines show slices for two fixed values of $\sqrt{\sigma_{\rm D}}/m_{\rm D}^3$ through the right panel of Fig.~\ref{fig:DMF}. For 
the parameters corresponding to the lighter solid line, the radius of the DM sphere never exceeds that of the NS, whereas in the
case marked by the darker solid line the crossing where the DM starts to envelope the NS is shown 
by the vertical dotted line. The horizontal dashed lines are for the upper bound $\Lambda = 800$~\cite{TheLIGOScientific:2017qsa}
and the tentative lower bound $\Lambda = 400$~\cite{Radice:2017lry}.}
\label{fig:MLH4}
\end{figure}

Similar results were recently obtained in Ref.~\cite{Nelson:2018xtr}, which focused on cases where a DM halo envelopes the NS. 
For the purpose of comparison, we shown in Fig.~\ref{fig:MLH4} two vertical
slices through the right panel of Fig.~\ref{fig:DMF} with fixed $\sqrt{\sigma_{\rm D}}/m_{\rm D}^3$.
We see that the tidal deformability depends essentially on the ratio of the DM mass contained within the NS
to that contained in the halo outside the NS.
In the case of the lighter solid line for $\sqrt{\sigma_{\rm D}}/m_{\rm D}^3 = 0.06$/GeV$^2$, the DM always forms a core
whose radius is smaller than that of the nuclear matter, and the tidal deformability decreases
monotonically as a function of the total DM mass.
On the other hand, in the case of 
the darker solid line for $\sqrt{\sigma_{\rm D}}/m_{\rm D}^3 = 16$/GeV$^2$,  after decreasing rapidly as the DM core radius
increases towards the NS radius, which it crosses when log$_{10} (M_{\rm D}/M_{\rm N}) \simeq -2.18$,
and decreasing more slowly for larger $M_{\rm D}/M_M$, $\Lambda$ starts to increase for log$_{10} (M_{\rm D}/M_{\rm N}) \gtrsim -1.5$.
We find that the minimum value of $\Lambda$ occurs for a DM halo radius larger than the NS, when 
the total DM mass inside the NS equals that outside the NS, and $\Lambda$ increases to above the case with no DM
when $M_{\rm D}/M_{\rm N}$ is sufficiently large.

%-------------------------------------------------------------------------------
\section{Conclusions}
\label{sec:conclusions}
%-------------------------------------------------------------------------------

We have presented in this paper a unified discussion of the possible effects of DM structures on the possible masses, radii and
tidal deformation parameters of a NS. Our main focus has been on DM cores with smaller radii than the nuclear matter,
though we have also considered some aspects of models in which the DM structure is a halo that envelopes the NS.
Qualitatively, we find that a DM core would tend to {\it decrease} the maximum NS mass, its radius for any given mass, and its
tidal deformability parameter $\Lambda$. These effects could be observable for DM cores with masses $\sim 5$\% of the
total NS mass, in which case the DM cores might also yield observable signatures in the frequency spectrum of GW emissions
following a NS-NS merger, as discussed in~\cite{Ellis:2017jgp}.

The DM effects on NS properties could complicate the interpretation of NS measurements in terms of possible models of
the NS EOS. For example, some EOS models that yield maximum masses $\gtrsim 2 M_\odot$ might no longer do so
if the NS had a heavy DM core. On the other hand, some EOS models that seemed to be incompatible with the upper limit
on the tidal deformation parameter $\Lambda < 800$ might be acceptable if the DM core were sufficiently massive.

We have also presented in the Appendix some specific scenarios for the formation of a DM core in a NS, which motivate our
calculations of the possible effects, though detailed explorations of these scenarios are beyond the scope
of this paper. However, one relevant comment is that the DM mass fraction would, in general, not be universal.
Scenarios that rely on mechanisms taking place inside the NS would give DM mass fractions that depend on its age or initial temperature,
while other scenarios could yield a DM mass fraction that depends on the environment in which it was formed.
Such possibilities would further complicate the interpretations of future GWs and other probes of NS properties.

In the mean time, we await eagerly future experimental measurements of NS properties that could cast light on DM, as well 
as the NS EOS and gravitational physics.

%-------------------------------------------------------------------------------
\acknowledgments
We would like to thank Kari J. Eskola, Jarkko Peuron, Hardi Veerm\"ae and Indrek Vurm for discussions. This work was supported by 
the Estonian Research Council grants MOBTT5, IUT23-6, IUT26-2, PUT1026, PUT799, PUT808, the ERDF center of Excellence project TK133,
the UK STFC grant ST/L000326/1 and the European Research Council grant NEO-NAT.

%-------------------------------------------------------------------------------
\appendix*

\hypertarget{sec:formation}{}
\section{Formation of Dark Matter Structures}
During its usual evolution, a NS cannot accumulate via gravitational accretion DM in quantities sufficient to form
substantial DM cores or halo. Assuming a NS lifetime of $\sim 10$\,Gyr and, for example,
a prototypical weakly interacting massive particle candidate, 
the typical amount of accreted DM does not significantly exceed 
$\sim 10^{-10} M_{\odot}$~\cite{Goldman:1989nd,Kouvaris:2007ay,Kouvaris:2010vv,Guver:2012ba}. 
Therefore the formation of substantial DM structures would require additional mechanisms involving the phenomenology 
of alternative DM models. Below, we provide three examples of such dynamics.

{\it Dark conversion of the neutron to scalar DM.} 
The lower limit on the lifetime of a bound neutron into invisible particles can be inferred from the bound $\tau(n \to 3 \nu) > 5 \times 10^{26}$~y
to be many orders of magnitude greater than the age of the Universe. However, neutrons inside a NS reach a Fermi momentum $p_F$
of several hundred MeV, which opens up possibilities for conversions of such Fermi neutrons to heavier DM particles $\chi$ 
with masses $\lsim m_n + {\cal O}(p_F^2/(2m_n))$ that are kinematically forbidden from being produced in the decays of normal bound neutrons.
Such a mechanism might enable a significant fraction of the nuclear matter inside the NS to be converted into DM particles. Notice that the stability of the ${}^9$Be nucleus requires $m_\chi>937.90$ MeV, implying that the DM particles produced in neutron decays will remain gravitationally bound to the NS. The NS escape 
velocity is $v_e = \sqrt{2GM_{\rm N}/R_{\rm N}}$. For $M_{\rm N}=1.4M_\odot$ and $R_{\rm N}=12\,{\rm km}$ 
the escape velocity is $v_e\simeq 0.6$, hence, given the available mass range, the DM particles will remain gravitationally bound to the NS.

As an example of a model in which such dark conversions of Fermi neutrons could form substantial DM cores or halos, 
we consider a scalar DM particle $\chi$ with an  effective Yukawa interaction 
\begin{equation}
	\mathcal{L} \supset y \bar{n} \chi \nu 
\end{equation}
with the neutron $n$ and a sterile neutrino $\nu$ that is assumed to have negligible mass. In such a scenario, the rate for the conversion process $n \to \chi$ is estimated on dimensional grounds to be
\begin{equation}
\Gamma_{n\to \chi} \sim y^2 E_F \,.
\end{equation}
where $E_F=m_n + p_F^2/(2m_n)$ is the Fermi energy. For a NS with a baryonic central energy density of $\mathcal{O}(10^{16}\,{\rm g}/{\rm cm}^3)$, we estimate the corresponding Fermi energy to be $E_F\sim 1.2\,{\rm GeV}$.  Assuming that $\Gamma_{n\to \chi}\ll t_{\rm NS}$, where $t_{\rm NS}$ is the age of the NS, we estimate the total DM mass fraction
produced in the NS to be  
\begin{equation}
\frac{M_{\rm D}}{M_{\rm N}} = \frac{m_\chi}{m_n} \Gamma_{n\to \chi} t_{\rm NS}\,.
\end{equation} 
Since the oldest NS have lifetimes of the order of $t_{\rm NS} \sim 10^{10}$~y, this mechanism could generate DM cores with $M_{\rm D} \simeq 0.05 \, M_{\rm N}$ if $y \sim 10^{-22}$. Consequently, DM and neutrons interact so feebly that the two sectors never achieve equilibrium.\footnote{{Dark decay of the neutron has been considered as a solution for the neutron lifetime puzzle. In that case the coupling has to be significantly larger. Consequently the baryonic and dark matter components reach chemical equilibrium inside a NS~\cite{Fornal:2018eol,McKeen:2018xwc,Cline:2018ami}, and our treatment does not apply.}} In this case, the two components should be treated separately when solving the TOV equations. Moreover, because the interaction strength we consider prevents additional NS cooling, we argue that the gravitational observables discussed previously and the possible presence of extra features in the frequency spectra of following the mergers of NS binaries~\cite{Ellis:2017jgp} are the only NS observables able to constrain this framework.

{\it Neutron bremsstrahlung of scalar DM.}
As an alternative to neutron conversions, light DM particles could be produced via bremsstrahlung in neutron-neutron scatterings, 
as proposed recently in Ref.~\cite{Nelson:2018xtr}. These processes convert kinetic energy to matter, hence this mechanism for 
DM production is more efficient in a younger and hotter NS. 

To exemplify the process, consider the effective coupling
\begin{equation}
	\mathcal{L} \supset g \bar{n} n \chi 
\end{equation}
between the neutrons and a scalar DM candidate $\chi$. This interaction produces
DM predominantly via the bremsstrahlung process $nn\to nn\chi$, which mainly proceeds via pion exchange. 
However, such reactions can yield a significant amount of DM only if the corresponding reaction rate is larger than that of 
usual NS cooling processes. Although a detailed study of DM production via bremsstrahlung lies beyond the scope
of this work, we give below a rough estimate to demonstrate that a DM structure with a mass of $O(1)\%$ of the NS mass
is within the reach of such a process.

Assuming an average neutron energy of $K_n = 3T/2$, where $T$ is the temperature
of the NS, DM production via bremsstrahlung is efficient only as long as $T>m_\chi/3$. 
Taking $v_e = 0.6$ as the escape velocity from a $1.4M_\odot$ NS, the produced DM particles would escape the 
NS if $T\gsim 0.5 m_\chi$. We can then estimate the temperature range relevant for the DM bremsstrahlung production as
\begin{equation}
m_\chi/3<T\lsim0.5m_\chi \,.
\end{equation}
A preliminary estimate of the maximal amount of DM produced is then given by comparing the average kinetic energy of neutrons inside the NS to their mass. 
The temperature of a newly born NS is $T_0\simeq 50\,{\rm MeV}$~\cite{Prakash:2000jr,Lattimer:2004pg}, 
corresponding to neutrons with an average kinetic energy of $K_0=3T_0/2\simeq 0.08m_n$. If $m_\chi = 100{\rm MeV}$, 
DM particles are produced via bremsstrahlung until $T=0.7T_0$, yielding a total DM mass
$M_{\rm D}\simeq (1-0.7)\times 0.08M_{\rm N} \approx 0.02 M_{\rm N}$. However, we point out that this estimate 
neglects the backreaction of the produced DM on the NS temperature, which might lead to an increase in its temperature and, therefore, 
to a larger total DM mass.

{\it Dark matter stars.}
A nonstandard DM sector with a subdominant dissipative component might lead to the formation of mixed stars 
(see, for instance,~\cite{Foot:2004pa,Fan:2013yva,Pollack:2014rja} and~\cite{Sandin:2008db,Leung:2011zz,Li:2012qf,Goldman:2013qla,Leung:2013pra,Mukhopadhyay:2016dsg,Panotopoulos:2017idn} for the NS case) characterized by a large DM mass fraction. Moreover, 
the asymmetric DM stars previously considered in~\cite{Kouvaris:2015rea,Maselli:2017vfi} could serve as the DM 
cores studied in the present paper. In fact, as demonstrated in~\cite{Pollack:2014rja}, a small fraction ($\lesssim 10\%$) of 
strongly self-interacting DM can lead to an early formation of black holes (BHs) that could serve as seeds for
subsequent supermassive BH formation, alleviating any tension with the standard picture. 
Depending on the shape of the initial perturbation spectrum, it is then plausible that BH formation be accompanied 
by the formation of smaller structures analogous to the NS of the visible sector. These compact dark objects could then 
serve as accretion centers for baryonic matter, potentially leading to the formation of a DM-admixed NS. 
This scenario does not require any interaction between the dark and the visible sectors, other than gravity, 
but a careful assessment of its viability would require dedicated studies.

\bibliography{core}
\end{document}